\documentclass[conference]{IEEEtran}
\IEEEoverridecommandlockouts
\usepackage{cite}
\usepackage{amsmath,amssymb,amsfonts}
\usepackage{algorithmic}
\usepackage{graphicx}
\usepackage{textcomp}
\usepackage{xcolor}
\usepackage{wrapfig}
\usepackage{textcomp}
\usepackage{xcolor}
\usepackage{soul}
\usepackage{braket}
\usepackage{algorithmic}
\usepackage[ruled,vlined, linesnumbered]{algorithm2e}
\usepackage{graphicx}
\usepackage{cite}
\usepackage{amsmath,amssymb,amsfonts}

\usepackage{graphicx}
\usepackage{textcomp}
\usepackage{xcolor}
\usepackage{wrapfig}
\usepackage{textcomp}
\usepackage{xcolor}
\usepackage{soul}
\usepackage{braket}
\usepackage{algorithmic}
\usepackage[ruled,vlined, linesnumbered]{algorithm2e}
\usepackage{graphicx}
\usepackage{cite}
\usepackage{amsmath,amssymb,amsfonts}

\usepackage{graphicx}
\usepackage{textcomp}
\usepackage{xcolor}
\def\BibTeX{{\rm B\kern-.05em{\sc i\kern-.025em b}\kern-.08em
    T\kern-.1667em\lower.7ex\hbox{E}\kern-.125emX}}
\begin{document}

\title{SHARE: Secure Hardware Allocation and Resource Efficiency in Quantum Systems\\
 }

\author{\IEEEauthorblockN{Suryansh Upadhyay}
\IEEEauthorblockA{\textit{The Pennsylvania State University} \\
\textit{University Park, PA, USA} \\
sju5079@psu.edu}
\and
\IEEEauthorblockN{Swaroop Ghosh}
\IEEEauthorblockA{\textit{The Pennsylvania State University} \\
\textit{University Park, PA, USA} \\
szg212@psu.edu}
}

\maketitle

\begin{abstract}
Quantum computing (QC) is poised to revolutionize problem-solving across various fields, with research suggesting that systems with over 50 qubits may achieve quantum advantage—surpassing supercomputers in certain optimization tasks. As the hardware size of Noisy Intermediate-Scale Quantum (NISQ) computers continues to grow, Multi-tenant computing (MTC) has emerged as a viable approach to enhance hardware utilization by allowing shared resource access across multiple quantum programs. However, MTC can also bring challenges and security concerns. This paper focuses on optimizing quantum hardware utilization in shared environments by implementing multi-programming strategies that not only enhance hardware utilization but also effectively manage associated risks like crosstalk and fault injection. We propose a novel partitioning and allocation method called Community-Based Dynamic Allocation Partitioning (COMDAP) and Secure COMDAP to refine and secure multi-programming capabilities in quantum systems. COMDAP ensures equitable and efficient resource distribution, addresses the issues of suboptimal partitioning, and significantly improves hardware utilization. We report a 23\% average improvement in hardware utilization rate compared to existing greedy heuristics, with rates averaging 92\%. COMDAP introduces an average increase of approximately 0.05X in $\Delta$CX, alongside a 3.5\% average reduction in PST across benchmarks.

\end{abstract}


\section{Introduction}

Quantum computing (QC) has drawn significant interest for its potential to transform problem-solving in numerous fields. Utilizing the unique principles of quantum mechanics, including superposition, entanglement and interference, these advanced computing systems can potentially achieve exponential speedup to tackle certain computational tasks compared with the classical computers. With potential uses that extend to machine learning\cite{b1}, security\cite{b2}, drug discovery\cite{b3}, and optimization\cite{b4}, quantum computing is becoming increasingly important to both academia and industry.

\begin{figure}
    \centering
    \includegraphics[width= 3.5in]{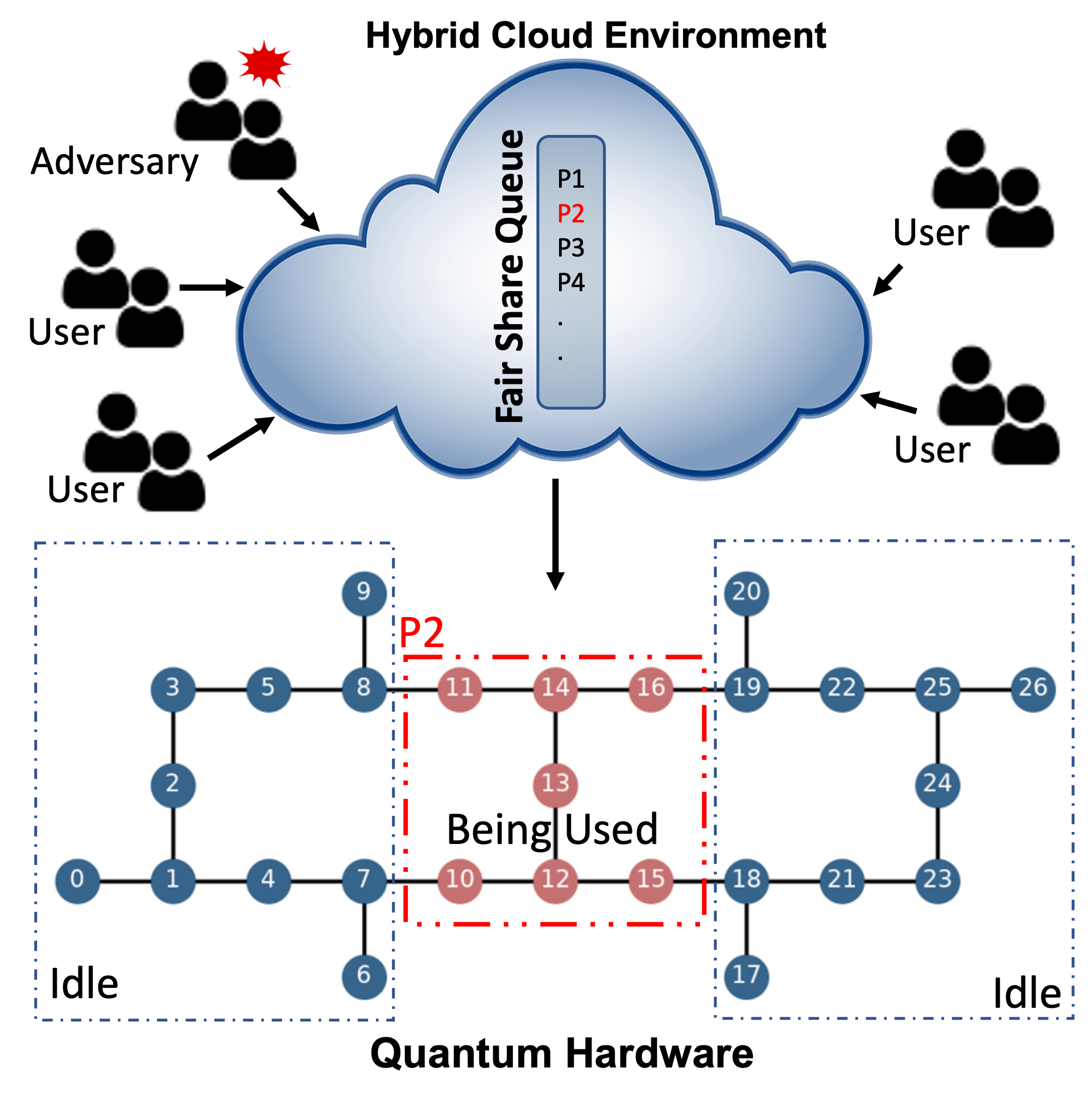}
    \caption{An exemplary quantum computing system operating at 25.93\% qubit utilization, executing a program for a single user while other users await access. This highlights need for multiprogramming approaches that enhance resource efficiency without compromising security against potential adversarial attacks such as fault injection and information stealth \cite{b9}\cite{b14}.}
    \label{1}
\vspace{-4mm}
\end{figure}

The practical implementation of quantum computing faces formidable challenges, including qubit decoherence, measurement errors, gate errors, and temporal variations. However, quantum computing continues to progress with the introduction of IBM's 133 qubit `Quantum Heron' which is distinguished by an architecture that significantly reduces error rates, improving upon the 127 qubit `Quantum Eagle'. Nevertheless, these devices fall into the Noisy Intermediate-Scale Quantum (NISQ) category and are limited by qubit connectivity and gate fidelity. Quantum computers are typically accessed via cloud services, which offer users convenience and scalability. Cloud-based quantum computing platforms from IBM and others offer remote access to quantum resources but face challenges such as job submission latency and queue backlogs, impacting efficient hardware utilization (Fig.\ref{1}). Addressing the efficient use of quantum resources without sacrificing circuit fidelity remains a critical concern in optimizing NISQ device performance. To address this challenge, the concept of multi-tenant computing (MTC) has emerged \cite{b6}\cite{b7}\cite{b8}\cite{b9}, gaining prominence with the rise in hardware size and improved qubit error rates.

\textbf{Motivation:} Multi-programming can significantly enhance the utilization of quantum hardware by mapping multiple quantum programs onto a single quantum hardware concurrently. However, this strategy may affect the reliability of programs due to several factors: a) limited availability of high-fidelity qubits, b) crosstalk from simultaneous gate operations, c) connectivity constraints necessitating SWAP operations for 2-qubit gates between non-adjacent qubits d) potential exploitation of connectivity constraints and crosstalk in multi-tenant environments for malicious purposes. Current partition methods, often based on greedy algorithms \cite{b6}, prioritize resource allocation to programs with high computational demands\cite{b6}\cite{b8}, typically those requiring a significant number of CNOT gates. This prioritization can result in inequitable treatment of less intensive programs, which may receive suboptimal qubits and mappings due to their lower priority. This approach can inadvertently provide adversaries opportunities to strategically exploit specific qubits or manipulate gaps in scheduling and partitioning policies \cite{b9}. Therefore, it is essential to develop robust and equitable resource allocation strategies that minimize interference to enhance both the reliability and security of quantum computations in multi-programming environments.

This paper seeks to answer a critical question: \textit{How can we efficiently and securely utilize quantum hardware in a shared environment?} We introduce a novel partitioning and allocation method that uses community detection algorithms to enhance the multi-programming capabilities of quantum hardware while enhancing the hardware-utilization and managing the associated security risks. Our approach is designed not just to create a single or few, highly optimized partition but to ensure that all partitions are fair and perform similarly. This approach also helps reduce the number of broken links—a common issue in poorly optimized partitions—and increases hardware utilization. By using the Louvain algorithm for community formation\cite{b10}, we initially identify natural groupings within a quantum system to form communities. To further refine the efficiency and effectiveness of partitioning, we introduce a new metric called the Connectivity and Reliability Index (CRI). This index evaluates the quality of partitions by considering the density of connections within communities and the error rates. By distributing quantum resources more equitably across the network, we aim to achieve a balance that not only enhances performance but also ensures fairness. This strategy is particularly effective in multi-tenant environments where equitable access to quantum resources is essential.

\emph{Our proposed framework builds on existing research and contributes to the field of quantum computing by providing innovative solutions for dynamic resource management.}

\textbf{Contributions:} In this work we (a) present a Greedy approach using the Connectivity and Reliability Index , (b) develop Community-Based Dynamic Allocation Partitioning (COMDAP), (c) enhance COMDAP with Secure COMDAP, (d) introduce the Connectivity and Reliability Index (CRI) to evaluate partition quality, (e) conduct exhaustive simulations using diverse quantum circuits and backends to validate the performance and efficiency of our partitioning heuristics.

\textbf{Paper organization:} Section II provides background information. Section III outlines the challenges in fair and secure resource allocation. In section IV we discuss our proposed heuristics. Section V presents simulations, results, and analysis. The discussions are presented in Section VI. Section VII concludes the paper.

\section{Background}

\subsection{Qubits and Quantum gates}

Qubits are similar to classical bits in that they store data through internal states such as $\ket{0}$ and $\ket{1}$. However, due to their quantum nature, qubits can exist in a superposition of both $\ket{0}$ and $\ket{1}$. A qubit's state, denoted by $\varphi$ = a $\ket{0}$ + b $\ket{1}$, can be expressed as a combination of complex probability amplitudes, $a$ and $b$, corresponding to the states $ket0$ and $ket1$. 
Mathematically, quantum gates are represented using unitary matrices (a matrix U is unitary if U$U^\dagger$ = I, where U$^\dagger$ is the adjoint of matrix U and I is the identity matrix). 

\subsection{Quantum errors}

Qubits are highly sensitive to noise and are error-prone. Key error types that affect quantum computing operations include:

\subsubsection{Coherence Errors} interactions between qubits and their surrounding environment lead to state decoherence, introducing errors that limit the effective circuit depth achievable on Noisy Intermediate-Scale Quantum (NISQ) devices \cite{b11}.

\subsubsection{Operational Errors} imperfections in quantum gate implementation, often caused by imprecise pulse delivery, result in operational errors. These errors manifest as inaccuracies in the rotational angles required for specific quantum operations, potentially causing qubits to under-rotate or over-rotate, thus deviating from their intended states \cite{b12}.

\subsubsection{Measurement Errors} flaws in measurement circuitry can lead to measurement errors, where qubits read as '0' may actually be '1', and vice versa. These readout errors can result in incorrect computational outcomes, regardless of the accuracy of preceding quantum operations \cite{b13}.

\subsubsection{Crosstalk} crosstalk between qubits during parallel gate operations presents a significant challenge, as it can degrade the performance of each involved qubit \cite{b14}. Crosstalk is particularly problematic in densely packed quantum circuits where multiple operations occur simultaneously.

\subsection{Quantum Cloud Services}

Quantum computers require extensive and expensive infrastructure, including cryogenic coolers and superconducting wires, making direct access challenging. Quantum cloud services like those offered by IBM, Google, and AWS Braket, provide users with remote access to the users worldwide while simplifying the management of these complex systems. However, several challenges persist:

\subsubsection{Resource Under-utilization} Current quantum cloud services execute only one program per run, leading to significant under-utilization of hardware. This operational mode leaves many qubits idle, especially in larger quantum systems, where the extent of unused resources can be substantial.

\subsubsection{Long wait times} Rapid increase in quantum computing research and demand outpaces the development/availability of new quantum computers, creating bottlenecks and extended wait times for access resulting in long wait queues. This is specifically a challenge for iterative algorithms.

\subsubsection{Security Risks} Access to quantum computing through cloud services like those offered by IBM, Google, and AWS Braket, while necessary, is expensive, may push users towards potentially untrustworthy third-party services. Such third-party services could pose risks by enabling the theft of sensitive intellectual property from quantum circuits, similar to risks associated with untrusted compilers \cite{b15}. 

\subsection{Multi-tenant computing and scheduler}

Multi-tenant computing (MTC) allows multiple programs or tasks to run concurrently on the same computer system. 
To ensure efficient and fair execution of quantum algorithms, it is critical to manage and optimize the allocation of quantum resources in a cloud-based environment. A scheduler is essential for orchestrating quantum jobs, assigning them to available quantum devices, and prioritizing them based on criteria such as \textit{Service-Level Agreements (SLAs), user priority and fairness, and resource constraints.}

\subsection{Fairness in Resource Allocation and Fair Share Queue}

In the context of this work, fairness involves the equitable allocation of quantum computing resources, particularly qubits and their connectivity, to each program operating on a shared quantum system. We assume that a fair share queue based on IBM's current policy\cite{b18} and a scheduler are already in place. When a quantum system is available for more tasks, it requests the next job from the fair-share scheduler. The scheduler typically selects the next job by first determining which group has used the least of their allocated time within the scheduling window. Within each group, program execution is prioritized based on the least used share, with the oldest jobs being prioritized first-in-first-out (FIFO). This policy  can also be used for MTC to select multiple programs for parallel execution on the quantum hardware. 

\subsection{Louvain community algorithm}

The Louvain Community Detection Algorithm is a widely-used method for identifying community structures within networks. It operates as a heuristic approach that optimizes modularity, which is a scale of how densely connected the nodes within a community are compared to nodes in different communities\cite{b10}. Initially, every node is assigned to its own community. The algorithm then examines each node to determine if there is a modularity gain by moving it to a neighboring community. For each node, it calculates the potential gain from moving the node to each of its adjacent communities. If moving a node results in a positive modularity gain, the node is relocated to the community with the maximum gain. If no community offers a positive gain, the node remains in its original community. The calculation of the modularity gain from moving an isolated node into a different community is essential for this process. The modularity gain \( \Delta Q \) can be computed with the following formula, derived from combining prior studies and algebraic adjustments:

\begin{equation}
\Delta Q = \left(\frac{\sum_{\text{in}} - (\sum_{\text{tot}} \times k_i)}{2m}\right) + \gamma \left(\frac{k_i \times k_{\text{in}}}{2m^2}\right)
\end{equation}

Here, \( \sum_{\text{in}} \) represents the sum of the weights of the links from nodes in the community to node \( i \), \( k_i \) is the sum of the weights of the links incident to node \( i \), \( \sum_{\text{tot}} \) is the sum of the weights of the links incident to all nodes in the community, \( m \) is the total weight of all edges in the graph, and \( \gamma \) is the resolution parameter, which adjusts the scale at which community detection operates.

\subsection{Relation to prior work}

The multi-programming environment for a qunatum hardware was first introduced by developing the Fair and Reliable Partitioning (FRP) method \cite{b7}. This method, which categorizes qubits based on utility derived from connectivity and error rates, often expands from a high utility root, leading to biased allocations favoring specific processor regions and overlooking crosstalk in hardware partitions. In contrast, our approach utilizes a community-based strategy with a community detection algorithm that identifies qubit groups by inter-connectivity and reliability. This not only captures clustering effects and local topologies but also ensures equitable resource distribution across the hardware. Unlike methods that focus on iterative expansion \cite{b6}\cite{b7} and prioritize individual qubit fidelity, our method optimizes the collective characteristics of qubit clusters to form robust sub-networks.

In contrast to the approach described in \cite{b5}, which employs reinforcement learning (RL) to optimize circuit placement by learning device noise characteristics, we utilize the Louvain algorithm for community detection. This method forms a hierarchy of partitions, diverging significantly from the RL-based strategy. Training an RL agent can be inherently complex and resource-intensive. In comparison the hierarchy tree is constructed once per calibration cycle and can be reused across multiple allocations without incurring additional computing overheads. In \cite{b8}, the authors present CDAP, a qubit mapping scheme that uses the Fast Newman (FN) algorithm to create initial mappings for concurrent quantum programs and establish a hierarchy tree. This method primarily focuses on minimizing crosstalk within partitions and overlooks the crosstalk between them, which introduces security vulnerabilities in multiprogramming environments. In contrast, our approach comprehensively mitigates these limitations by developing equitable and secure partitions that address security threats including crosstalk and adversarial SWAP injection.

\section{Challenges in Fair and Secure Resource Allocation}

In our work, we address the following challenges to boost throughput, ensure fair resource distribution and minimize security risks for simultaneous execution of multiple programs on NISQ computers:

\subsection{Qubit variability}

The varying error rates and coherence times of qubits present substantial challenges for fair resource allocation in multi-programming environments. These error rates are not constant but fluctuate over time, which directly affects the reliability of the computations that the allocated qubits perform \cite{b16}. For example, consider a five-qubit quantum system running two programs: Program1 needs 3 qubits and Program2 needs 2. Qubits 1, 2, 3, and 5 have lower error rates compared to the more error-prone qubit 4. When run separately, Program1 uses qubits 1, 2, and 3, while Program2 would typically use a subset of these. However, when both programs run simultaneously, there's a compromise due to limited qubit availability. Program1 still uses qubits 1, 2, and 3, but Program2 has to use qubits 4 and 5, despite qubit 4's higher error rate.

\subsection{Connectivity constraint and SWAP operations}

Application reliability in quantum computing is influenced by qubit allocation, program characteristics, and the network topology of the assigned region. In superconducting quantum systems, qubits are interconnected via resonators (waveguides), which are integral for implementing multi-qubit gates such as the controlled-NOT (CNOT) and controlled-Z (CZ) gates used in IBM and Rigetti quantum systems, respectively. Not all qubits are interconnected and this restricted connectivity poses a challenge in quantum circuit mapping (known as coupling constraints), which is addressed by routing qubits using the SWAP operation to ensure that logical qubits requiring 2-qubit operations are in close proximity, albeit at the expense of:

\textbf{a) Increase in circuit depth:} extra SWAP operations result in an increase in the overall depth of the quantum circuit.

\textbf{b) Computational overhead:} each SWAP operation consumes additional gate resources and prolongs the execution time of the circuit.

\textbf{c) Error accumulation:} as each SWAP operation introduces its own sources of error, extra SWAP operations contribute to error accumulation.

A well-connected topology can reduce the number of SWAP operations needed to position non-adjacent qubits for executing a CNOT gate.

\subsection{Security}

\subsubsection{Crosstalk Induced Faults} Crosstalk represents a significant security risk, particularly as quantum computers are expected to operate in multi-user environments, providing services via public cloud platforms. This shared environment allows for the operations of neighboring circuits to interfere with each other through crosstalk. Such interference can be intentionally leveraged to inject fault, resulting in degraded program performance or incorrect outputs \cite{b14}.

\subsubsection{Adversarial SWAP Injection} Quantum jobs, when submitted to a quantum system, are queued in the scheduler alongside other submissions, awaiting execution. This scheduling process, which determines the execution sequence and concurrency of tasks, can be exploited due to the connectivity constraints and \textit{unfair priority based hardware allocation} approaches. By strategically occupying specific qubits, an adversary can force a significant increase in SWAP operations in the victim’s program \cite{b9}.

\begin{figure*}
    \centering
    \includegraphics[width= 7.15in]{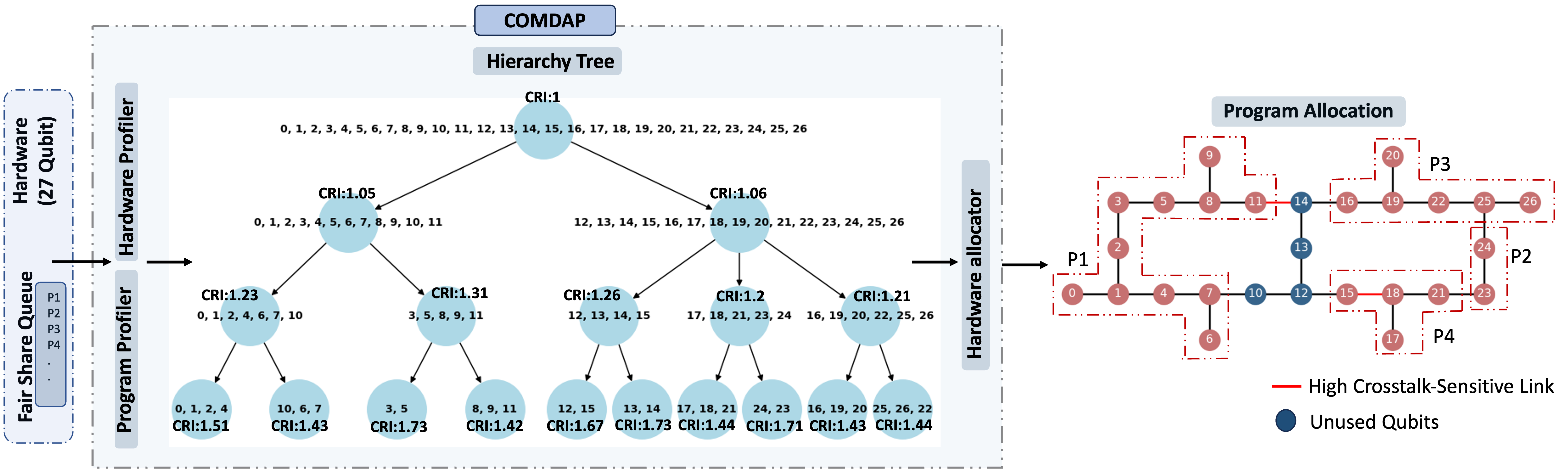}
    \caption{Proposed COMDAP framework for allocating quantum programs to hardware. Utilizing the Louvain algorithm, COMDAP structures hardware into a hierarchy tree where communities are identified based qubit connectivity and CNOT error rates. Program allocation is then performed by assessing communities with the CRI metric, ensuring optimal and feasible partitioning for queued programs.}
    \label{2}
\vspace{-4mm}
\end{figure*}

\section{Fair and Secure Hardware Allocation}

\subsection{Assumptions and approach}
In this work, we assume that a fair share queue and a scheduler are already in place. We propose and benchmark four distinct hardware partitioning heuristics: the Greedy heuristic using an attractor node approach (which serves as our baseline), Greedy Hardware Allocation using CRI, COMDAP: Community-Based Dynamic Allocation Partitioning, and Secure COMDAP that incorporates crosstalk considerations. Additionally, we introduce the Connectivity and Reliability Index (CRI), a metric we employ to quantify and enhance the efficiency and effectiveness of our partitioning methods.

\subsection{Connectivity and Reliability Index (CRI)}

This metric is designed to quantitatively assess and compare partitions within quantum hardware. Traditional methods of comparing partitions often fall short due to disparities in relative partition sizes, making direct comparisons unfeasible. To address this, the CRI provides a normalized metric that evaluates each partition based on its connectivity and error performance relative to the original, unpartitioned hardware. Specifically, the CRI measures how good/bad a given partition is compared to the unpartitioned hardware. By doing so, it provides a standardized index and allows for more accurate and fair comparisons between different hardware partitions, ensuring that each is assessed on a like-for-like basis with regard to its connectivity and reliability. \textit{A CRI value of 1 indicates that the partition's performance is equivalent to that of the original hardware. If the CRI exceeds 1, the partition is considered to outperform the original hardware in terms of connectivity and error rates. Conversely, a CRI value less than 1 suggests that the partition's performance is inferior to that of the original hardware}.
Mathematically, it is defined as follows:

\begin{equation}
CRI = \frac{\frac{D_{\text{partition}}}{\text{C}_{\text{partition}}} + \alpha \left(1 - \left(E_{\text{partition}} + R_{\text{partition}}\right)\right)}{\frac{D_{\text{Hardware}}}{C_{\text{Hardware}}} + \alpha \left(1 - \left(E_{\text{Hardware}} + R_{\text{Hardware}}\right)\right)}
\end{equation}

where:

a) Density (D): is defined as the ratio of the total number of actual edges to the total number of possible edges in the graph. Mathematically, it can be expressed as:

\begin{equation}
D = \frac{2 * E}{N \times (N - 1)}
\end{equation}

where \(E\) represents the number of actual edges, and \(N\) is the number of nodes in the graph. A higher density value indicates a greater degree of connectivity among the nodes, suggesting more robust communication or interaction capabilities within the partition/hardware.

b) Compactness (C): evaluates the degree to which elements within a partition or the entire quantum hardware are closely grouped or interconnected. A lower compactness score indicates that the elements are more densely packed. Compactness is quantitatively defined as:
\begin{equation}
C = \frac{\text{diameter}}{\text{max possible diameter}}
\end{equation}

where the \textit{diameter} represents the maximum distance between any two nodes in coupling map of that partition or hardware.

c) Alpha ($\alpha$): is a weighting factor used to balance the influence of error terms against connectivity and compactness measures. Adjusting $\alpha$ allows for emphasizing either structural or error aspects depending on the specific requirements or goals of the partitioning strategy. For all our evaluations we set $\alpha$ to 1.

d) Average CNOT error (E): represents the average CNOT error rate of a partition/hardware. It quantifies the operational errors associated with CNOT gate operations.

e) Average readout error (R): represents the readout error rate of a partition/hardware.

\subsection{Attractor Node-Based Greedy Hardware Allocation}
The Greedy attractor node based heuristic, similar to prior works \cite{b6}\cite{b7}, identifies optimal nodes—attractor nodes—for constructing hardware partitions. These nodes are selected based on their connectivity and error rates, which are quantified using the Composite Fidelity Metric (CFM):
\vspace{-.25mm}

\begin{equation}
CFM = d + \left(1 - (avg\_cnot + r_e)\right)
\end{equation}

\begin{algorithm}[t]
\caption{Greedy Hardware Allocation}
\label{alg:greedy_allocation}
\SetAlgoLined
\KwIn{$p$: program\_properties, $G$: hardware graph}
\KwOut{Allocated partitions and hardware utilization}

\textit{allocated\_partitions} $\gets$ []\\
\textit{allocated\_nodes} $\gets$ set()\\

\For{\textit{size} in \textit{program\_sizes}}{
    \textit{max\_metric} $\gets$ $-\infty$\\
    \textit{best\_partition} $\gets$ None\\
    
    \For{\textit{subgraph} in possible\_subgraphs}{
        \textit{current\_metric} $\gets$ calculate\_metric\\
        \If{\textit{current\_metric} $>$ \textit{max\_metric}}{
            \textit{max\_metric} $\gets$ \textit{current\_metric}\\
            \textit{best\_partition} $\gets$ \textit{subgraph}\\
        }
    }
    
    \If{\textit{best\_partition}}{
        \textit{allocated\_partitions}.append((\textit{best\_partition}, \textit{max\_metric}))\\
        \textit{allocated\_nodes}.update(\textit{best\_partition}.nodes())\\
        }
}

\KwOut{\textit{allocated\_partitions}, \textit{hardware\_utilization}}
\end{algorithm}

Where, \(d\) represents the degree of the qubit, \(avg\_cnot\) is the average CNOT error, and \(r_e\) denotes the readout error. The CFM effectively integrates the node's connectivity (degree) with penalties for average CNOT and readout errors. Nodes that achieve higher scores are preferred for their robust connectivity and lower error rates, making them ideal for forming subgraphs tailored to specific quantum programs. Partition formation begins with a breadth-first search (BFS) from a selected attractor node. This approach involves exploring neighboring nodes that are neither part of any existing subgraph nor previously allocated. The selection of candidates is based on their CFM scores, prioritizing those with the best metrics. This process is continued until the subgraph either meets the required size or exhausts all potential nodes. Once a subgraph reaches the necessary size, it is assessed by summing the CFM scores of its constituent nodes. The subgraph with the highest total score is then chosen as the optimal partition for a program.

\subsection{Greedy Hardware Allocation using CRI}

The Greedy Hardware Allocation Strategy, outlined in Algorithm \ref{alg:greedy_allocation}, is designed to optimally distribute hardware resources across quantum programs in the queue as per their specific requirements (size, qubits and connectivity). It iteratively selects the most appropriate hardware partition for each program, aiming to maximize the CRI.

First, the heuristic initializes an empty set to track nodes allocated to any program, ensuring no double-booking of resources. It then examines all viable subgraphs of the quantum hardware graph G (generated using the coupling map of the hardware), excluding any previously allocated nodes, to identify potential partitions for each program based on size requirements. For every candidate subgraph, the algorithm computes CRI. The subgraph with the highest CRI is chosen as the best partition for that program. Nodes within the selected partition are then marked as allocated, and the partition's details are stored. If no appropriate partition is found, an empty graph with a null metric is recorded, signaling no allocation for that program. This process repeats for each program in the queue until all have been assigned partitions or all the viable qubit/qubit-links have been used.

\begin{algorithm}[t]
\caption{COMDAP}
\label{alg:quantum_allocation}
\SetAlgoLined
\KwIn{$p:$ program\_properties, $G:$ hardware graph}
\KwOut{Allocated partitions and hardware utilization}

\For{each $program$ in $queue$}{
    \textit{partition\_found} $\gets$ False\\
    \For{each $community$ in hierarchy tree}{
        \If{$community$ size $\geq$ $program$ size }{
            \textit{subgraph} $\gets$ $community$\;
            {
                \textit{CRI} $\gets$ \texttt{calculate\_CRI(subgraph)}\;
                \If{$community$ size == $program$ size and \textit{CRI} is the highest}{
                    Allocate $community$ to $program$\;
                    Update \textit{hierarchy tree}\;
                    \textit{partition\_found} $\gets$ True\;
                    Break;  
                }
                \If{no exact match and \textit{CRI} highest for largest $community$}{
                    Allocate densest subset to $program$\;
                    Update \textit{hierarchy tree}\;
                    \textit{partition\_found} $\gets$ True\;
                    Break;  
                }
            }
        }
    }
    \If{not \textit{partition\_found}}{
        // Merge smaller communities\;
        \textit{base\_community} $\gets$ closest size match with highest CRI\;
        \textit{closest\_community} $\gets$ nearest smaller community for merging\;
        Merge \textit{base\_community} and \textit{closest\_community}, recalculate CRI\;
        Allocate $community$ to $program$\;
        Update \textit{hierarchy tree}\;
    }
}

\KwRet{\textit{allocated\_partitions}, \textit{hardware\_utilization}}
\end{algorithm}

\begin{figure*}
    \centering
    \includegraphics[width= 7.15in]{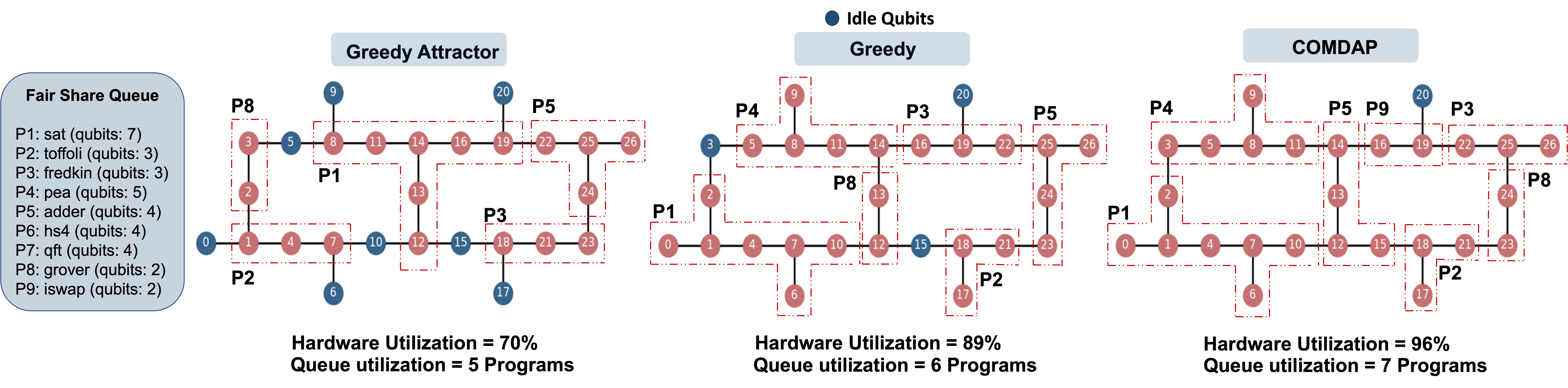}
    \caption{Comparison of hardware partition strategies: On the left, a representative fair share queue with 9 programs. The greedy attractor heuristic allocates only 5 programs, leaving many qubits isolated. The greedy heuristic using CRI, shown in the middle, performs better. COMDAP achieves maximum hardware and queue utilization.}
    \label{3}
\vspace{-4mm}
\end{figure*}

\subsection{Community-Based Dynamic Allocation Partitioning (COMDAP)}

The COMDAP program allocation strategy is outlined in Algorithm \ref{alg:quantum_allocation}. It utilizes the Louvain algorithm to optimally partition a quantum hardware. COMDAP begins with a graph derived from the hardware's coupling map, where CNOT error rates are represented as edge weights. The Louvain Community Detection Algorithm then identifies highly cohesive communities within this graph, employing a heuristic approach based on modularity optimization. This ensures that every partition is both feasible and optimally configured. The COMDAP allocation strategy is structured into two phases:

\subsubsection{Hardware profiling and hierarchy tree}

A critical component of our allocation strategy is the construction of a hierarchy tree, based on the Louvain community detection algorithm Fig.\ref{2}. This tree has the following properties:

\textbf{a) Node Representation:} Each node within the hierarchy tree represents a community of tightly interconnected physical qubits, making it a potential candidate set for initial allocation.

\textbf{b) Interconnection and Reliability:} Nodes consist of physical qubits chosen for their low CNOT error rates and robust connectivity, ensuring a high degree of reliability.


\textbf{c) Efficiency Across Calibration Cycles:} Constructed once per calibration cycle, the hierarchy tree can be reused without additional computing overheads.

\textbf{d) Resource Identification and Utilization:} The tree structure aids in superior initial mapping by structuring physical qubits based on their reliability and connectivity avoiding the pitfalls of non-strategic, greedy algorithms.

\subsubsection{Hardware Allocation for Queued Programs}

The allocation of quantum programs to physical qubits is achieved by systematically exploring the hierarchy tree from the bottom up. This tree structures the quantum hardware graph into communities, each evaluated and characterised using the CRI metric which assesses each community based on factors such as average path length, node connectivity, density and error rates.

The process of matching quantum programs to communities considers several scenarios, depending on the available communities and their characteristics:

a) If a community exactly matching the program's size and possessing the highest CRI is found, it is directly allocated to that program.

b) If no exact match is available, the heuristic identifies the largest available community with the highest CRI. It then selects the most densely connected subset within this community that meets the program's size requirements and has the highest CRI.

c) If there is no larger community available (excluding the complete hardware as a community), the strategy involves combining smaller communities. The heuristic first finds a base community that closely matches the program size requirements and has the highest CRI. It then searches for the nearest smaller community to merge, aiming to maximize the CRI for the resultant partition.
    
This method ensures an equitable resource distribution across programs, making each partition as close to optimal as possible given the available hardware and program constraints.

\begin{figure}
    \centering
    \includegraphics[width= 3.5in]{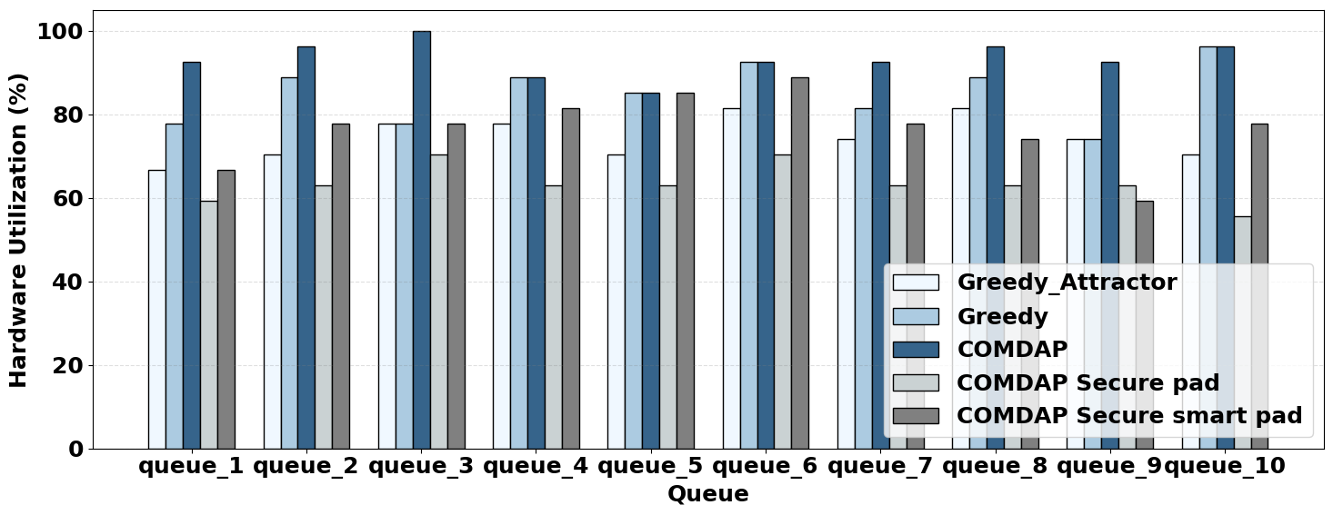}
    \caption{Comparison of hardware utilization for various partitioning methods on the Fake27QPulseV1 backend across ten different queues.}
    \label{4}
\vspace{-4mm}
\end{figure}

\subsection{Secure COMDAP}

Secure COMDAP enhances the original COMDAP framework by integrating heuristics designed to mitigate crosstalk, which is essential for effectively executing parallel quantum programs securely and with high fidelity. This section details the methodologies and enhancements introduced in secure COMDAP, including crosstalk characterization and the implementation of targeted padding strategies.

\subsubsection{Crosstalk Characterization} In Secure COMDAP, we leverage well-established research to address crosstalk in quantum computing environments effectively. Drawing on methodologies from references \cite{b19} and \cite{b20}, we focus on significant crosstalk effects observed at one-hop distances between CNOT pairs such as P$_i$($CX_{0,1}$) and P$_j$($CX_{2,3}$). Findings from \cite{b20} indicate that the pairs prone to strong crosstalk remain consistent over time, thus reducing the need for continuous re-characterization. Using results from Simultaneous Randomized Benchmarking (SRB)\cite{b20}, we note that when two qubits pairs, P$_i$ and P$_j$, operate concurrently, the error rates E(P$_i$-P$_j$) and E(P$_j$-P$_i$) significantly exceed those of solo operations. 

\subsubsection{Heuristic1-Program Padding} Program Padding involves allocating buffer qubits around the active community used by a program. These qubits are marked as occupied, and hence are unavailable for use in other programs. This setup forms a physical isolation layer that effectively mitigates the risk of crosstalk by preventing the simultaneous operation of CNOT pairs across different programs.

\subsubsection{Heuristic2-Smart Padding} Smart Padding refines the basic concept of program padding by selectively applying it based on detailed crosstalk characterization data. This method specifically targets qubit links identified as highly susceptible to crosstalk. We implement a threshold criterion from \cite{b20}, considering crosstalk significant if the correlated CNOT error rate exceeds three times the baseline error (E(P$_i$-P$_j$) $>$ $3 *$ E(P$_i$)). This selective approach allows us to concentrate our mitigation efforts on the most critically affected qubit pairs. If 2 qubit pairs (P$_i$, P$_j$) are identified with a high correlated error rate, smart padding is used to block the use of adjacent qubits in concurrent operations, thereby drastically reducing crosstalk potential. For example, in Fig. \ref{2} qubit pairs (q8,q11) and (q14,q16) are identified with highly crosstalk-sensitive links. If a partition includes the pair (q8,q11), we pad the directly connected qubit q14 to prevent any other program from including the pair (q14,q16). This effectively reduces crosstalk potential by ensuring these pairs do not overlap in parallel-running programs. However, if pairs with such links exist within the same partition, no padding is applied, since the risk of crosstalk-induced adversarial fault injection mainly concerns parallel executions. This ensures that padding is judiciously used only to mitigate crosstalk between separate programs.

\begin{figure}
    \centering
    \includegraphics[width= 3.5in]{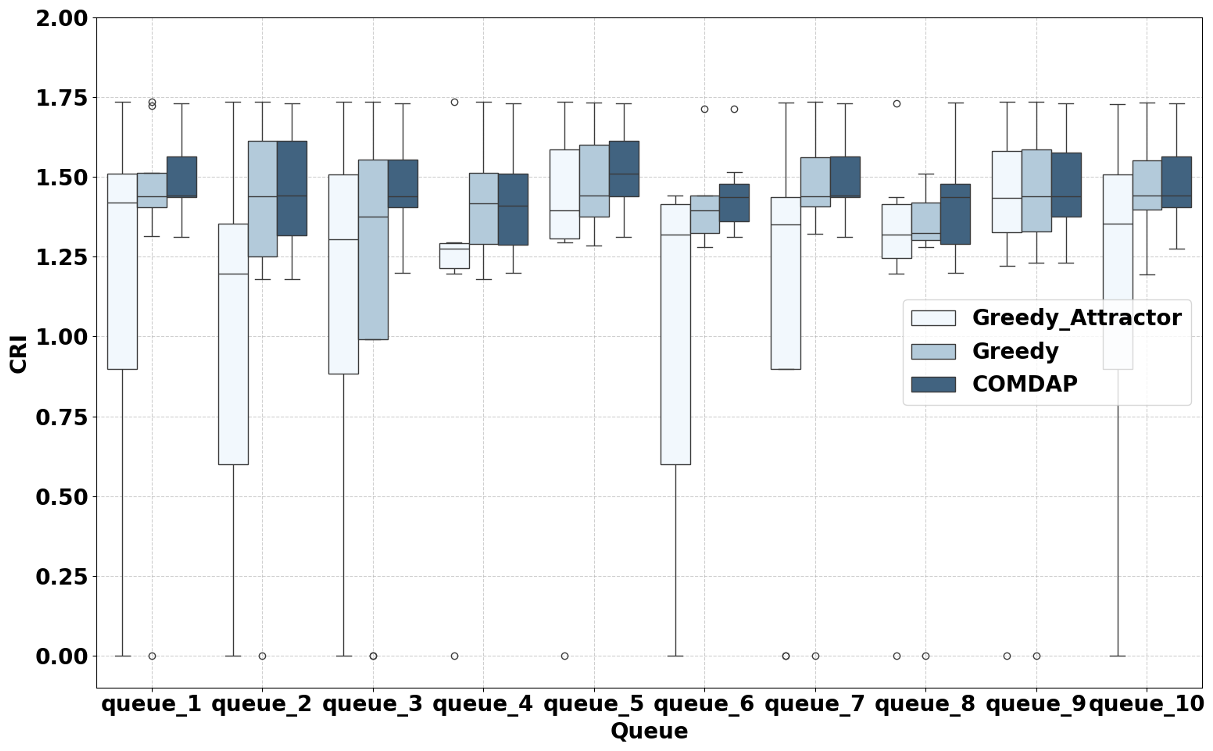}
    \caption{ Distribution of Community Reliability Index (CRI) values across ten queues for partitions allocated using COMDAP, Greedy heuristic, and Greedy Attractor-node strategies.}
    \label{5}
\vspace{-4mm}
\end{figure}

\section{Results and Analysis}

\subsection{Experimental setup}

We utilize the Qiskit open-source quantum software development kit for simulating quantum environments. For \textbf{benchmarking}, we use a variety of quantum circuits: iswap\_n2, hs4\_n4, simon\_n6, linearsolver\_n3, grover\_n2, qec\_en\_n5, toffoli\_n3, adder\_n4, qec\_sm\_n5, inverseqft\_n4, pea\_n5, fredkin\_n3, basis\_trotter\_n4, adder\_n10, basis\_change\_n3 among others, sourced from \cite{b21}. These benchmarks encompass a range of qubit counts (2-12), gate counts (4-1000), circuit depths, and connectivity patterns. To evaluate our approach, we form queues from these benchmarks, with queue depths surpassing the number of available hardware qubits, and containing programs of varying sizes and priorities.  For \textbf{benchmark execution}, we primarily utilize Qiskit's fake provider module Fake27QPulseV1 (27 qubit noisy simulator mimicking IBM's Hanoi system). We also test our heuristics across multiple fake provider modules, each with different configurations and size. These include FakeCambridge(28 qubits), FakeJohannesburg(20 qubits), FakeMelbourne(14 qubits), FakeRueschlikon(16 qubits), and FakeSingapore(20 qubits).

\begin{figure}
    \centering
    \includegraphics[width= 3.5in]{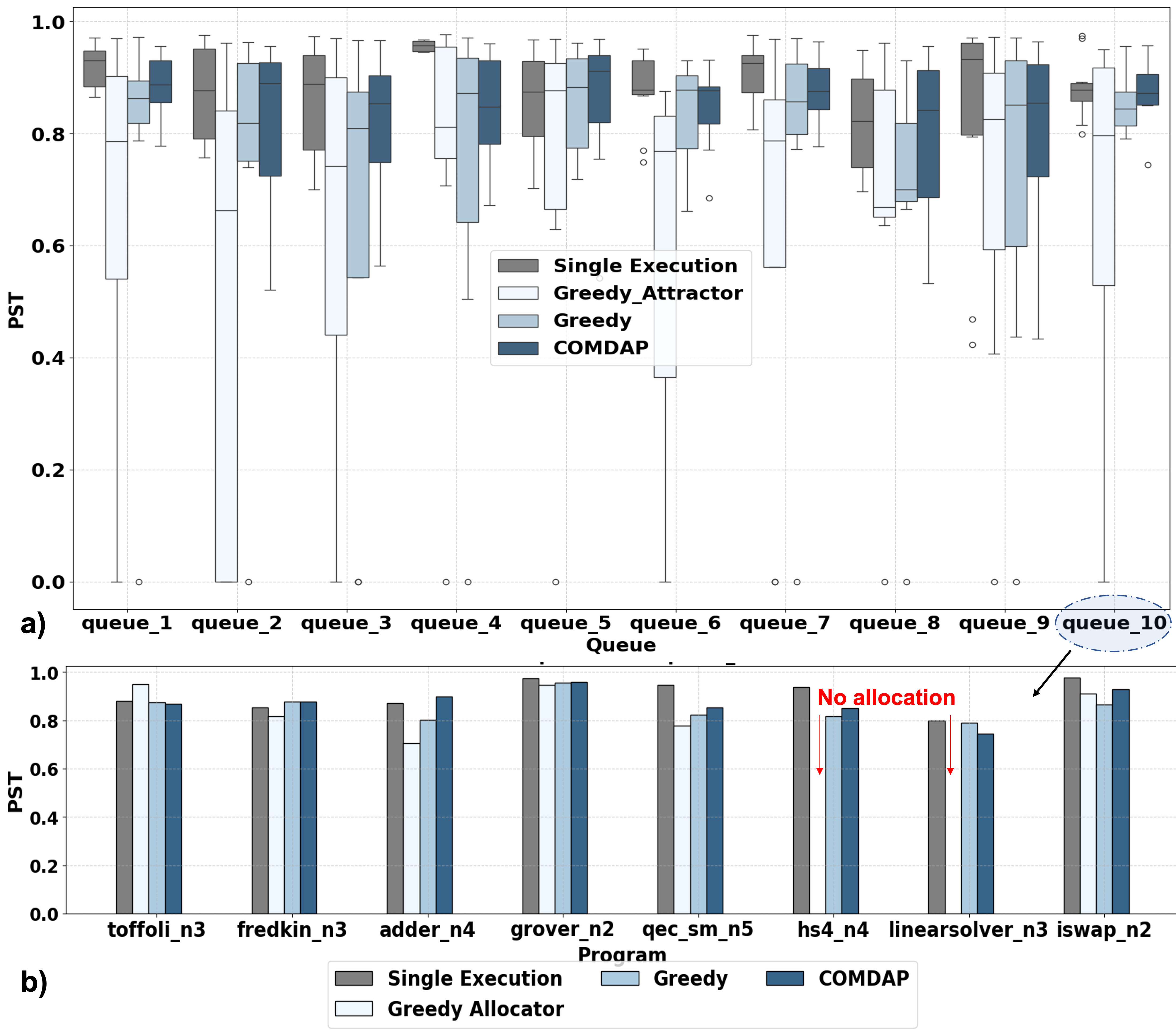}
    \caption{a) Comparative data on the impact of different hardware partitioning heuristics on PST in a multi-tenant computing environment. The baseline for comparison is single execution, which represents data from programs run individually. b) PST variation for programs in queue 10. }
    \label{6}
\vspace{-4mm}
\end{figure}

\subsection{Performance metrics}

We evaluate our framework's effectiveness using following key performance metrics:

\textbf{1)Probability of a Successful Trial (PST):} Defined as the fraction of trials that yield a correct result. A higher PST indicates superior fidelity.

\textbf{2)Number of Post-Mapping CNOT Gates:} We evaluate the efficiency of our partitioning policy by evaluating the number of CNOT gates required post-mapping, which reflects SWAP overheads.
    
\textbf{3)Post-Mapping Circuit Depth:} We also assess the depth of the circuit following partition allocation.
    
\textbf{4)Connectivity and Reliability Index (CRI):} It evaluates partition quality; a higher CRI indicates better connectivity and performance.

\subsection{Time complexity and Execution time}
We evaluate the time complexities of the proposed heuristics for hardware partitioning—Greedy Attractor Node-Based, Greedy CRI-Based, and COMPDAP—and note substantial differences in scalability and efficiency. The \textbf{Greedy CRI-Based heuristic}, which identifies connected subgraphs in a graph \( G \), faces potentially exponential time complexity \( O(N^{\text{target\_size}}) \), due to its recursive depth-first search that extends subgraphs by adding neighboring nodes. It can lead to a combinatorial explosion, especially in dense graphs. The \textbf{Greedy Attractor Node-Based heuristic} sorts physical and logical qubits and merges high-fidelity neighbor qubits into sub-partitions, resulting in a polynomial time complexity \( O(mk^2 + n \log n ) \) where \( m \) is the number of starting points, \( k \) the size of sub-partitions, and \( n \) the total nodes, making it more scalable. \textbf{COMPDAP} leverages a community detection approach with a logarithmic complexity \( O(n \log n) \), making it efficient for partitioning larger hardware setups. Moreover, for COMPDAP, the hierarchy tree is constructed once per calibration cycle and can be reused without additional computing overhead incurring negligible time overhead. We report an average execution time per queue of \textit{5 s} for the Greedy CRI-Based algorithm, \textit{10 ms} for the Greedy Attractor Node-Based algorithm, and \textit{2 ms} for the COMPDAP algorithm. While \textit{the Greedy CRI-Based algorithm struggles with scalability and is time-intensive for larger graphs or higher target sizes, the COMPDAP heuristic provides a more scalable and time-efficient solution for hardware partitioning in multi-tenant computing environments.}

\begin{figure}
    \centering
    \includegraphics[width= 3.5in]{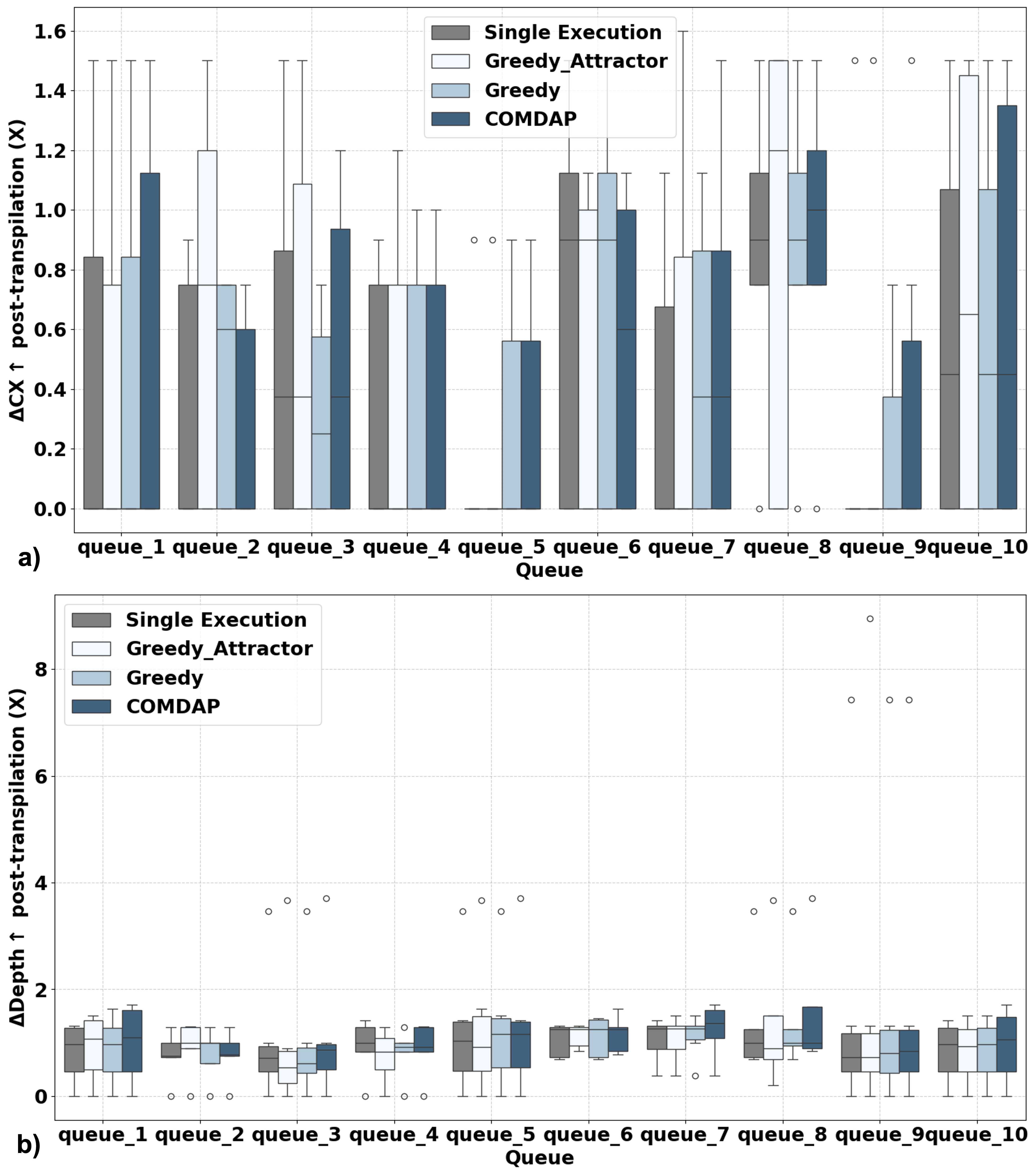}
    \caption{Comparison of a) SWAP overhead and b) change in circuit depth across different heuristic approaches.}
    \label{7}
\vspace{-4mm}
\end{figure}

\subsection{Hardware Utilization}
Fig. \ref{4} illustrates hardware utilization for various partitioning heuristics presented in this paper, tested across ten different varieties of queues on the fake-backend Fake27QPulseV1. For the attractor node-based greedy strategy, similar to \cite{b6}, we observe the lowest hardware utilization, with a maximum of 82\% and an average of 75\%. In contrast, our CRI-based greedy heuristic achieves a higher average utilization of 82\% with a max of 95\%. The COMDAP heuristic is the most effective, achieving 100\% utilization for a given queue, with an overall average of 92\%. Secure COMDAP, which includes a general padding approach to reduce interprogram crosstalk, exhibits a lower average utilization of 62\%. However, secure COMDAP with a smart padding strategy has an average 78\% utilization while effectively addressing crosstalk issues. \textit{Unlike attractor node-based greedy strategy that optimize for individual partitions, COMDAP ensures that all partitions are fair and perform similarly, significantly reducing broken links and enhancing overall hardware utilization} Fig. \ref{3}.

\subsection{Partition Quality}
We evaluate partition quality and allocation fairness using CRI. A value of 1 indicates a partition's performance matches the original hardware, while values above 1 indicate better relative connectivity and lower error rates, suggesting superior performance. Using our COMDAP heuristic on a 27-qubit hardware , all communities formed have a CRI over 1, indicating better performance compared to the baseline hardware Fig. \ref{2}. Notably, smaller communities show higher CRI values due to their denser connections. Fig. \ref{5} illustrates the distribution of CRI values for programs allocated partition in a queue. The CRI values are calculated only for programs in a queue that have received an allocated partition through COMDAP. An outlier value of 0 indicates no allocation. We observe that COMDAP consistently exhibits a higher median CRI across different queues, suggesting superior partition quality compared to the CRI-based Greedy heuristic and the Greed Attractor-node strategy. Additionally, \textit{COMDAP demonstrates less variability in CRI values, indicating a more equitable partition quality across programs of varying priorities}.

\begin{figure}
    \centering
    \includegraphics[width= 3.5in]{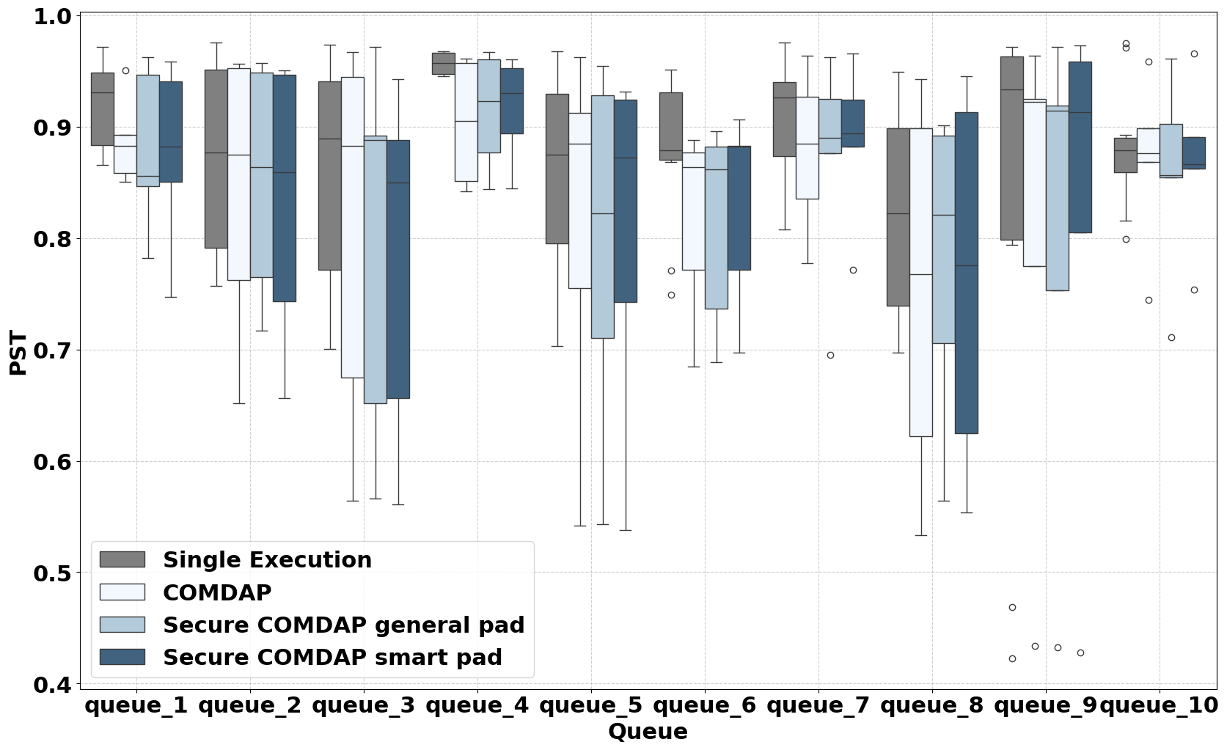}
    \caption{Comparative data on the impact of Secure COMDAP on PST in a multi-tenant computing environment. The baseline for comparison is single program execution. }
    \label{8}
\vspace{-4mm}
\end{figure}

\subsection{Fidelity and PST}
In Fig. \ref{6}(a), we report comparative data on the impact of various hardware partitioning heuristics on PST in a multi-tenant computing environment. The baseline for our analysis is independent execution, wherein each program is run separately. We observe an average PST reduction across queues by 13\%, 6\%, and 3.5\% for the heuristics: Greedy based on attractor node, Greedy using CRI, and COMDAP respectively.

We observe that the Greedy based heuristics may initially outperform COMDAP by effectively allocating high-priority programs through locally optimal qubit partitions, leading to improved PST for these programs Fig.\ref{6}(b). However, as these strategies progress through the queue, the quality of the remaining qubits/qubit connectivity likely declines due to fragmentation or high error rates, resulting in poorer PST. In contrast, the COMDAP tends to maintain a relatively stable PST for all programs suggesting that \textit{it employs a more equitable partitioning approach, ensuring a fair distribution of resources among all programs}.

\subsection{Program depth and Swap overhead}

We evaluate the impact of proposed hardware partitioning heuristics on the post-transpilation increase in the number of CNOT gates ($\Delta$CX), serving as an indicator of SWAP overhead, and the variation in the quantum circuit's depth post-transpilation ($\Delta$depth), which influences the circuit's execution duration and potential error incidence. The benchmark for this comparison is the single program execution, wherein programs are executed individually. We compare data across ten distinct queues, ensuring that only programs with an allocated partition shared by all three of our heuristics were included in the analysis to avoid data bias.

We observe a modest performance degradation with the Greedy attractor node-based heuristic, showing an average $\approx$0.1X increase in $\Delta$CX compared to single execution across all queues (Fig. \ref{7}(a)). In contrast, COMDAP exhibited a nominal average increase of $\approx$0.05X in $\Delta$CX. Meanwhile, the CRI-based greedy heuristic demonstrated no change in average delta CX. Additionally, both COMDAP and the CRI-based greedy heuristic showed a slight average depth increase of $\approx$0.1X (Fig. \ref{7}(b)).

\begin{figure}
    \centering
    \includegraphics[width= 3.5in]{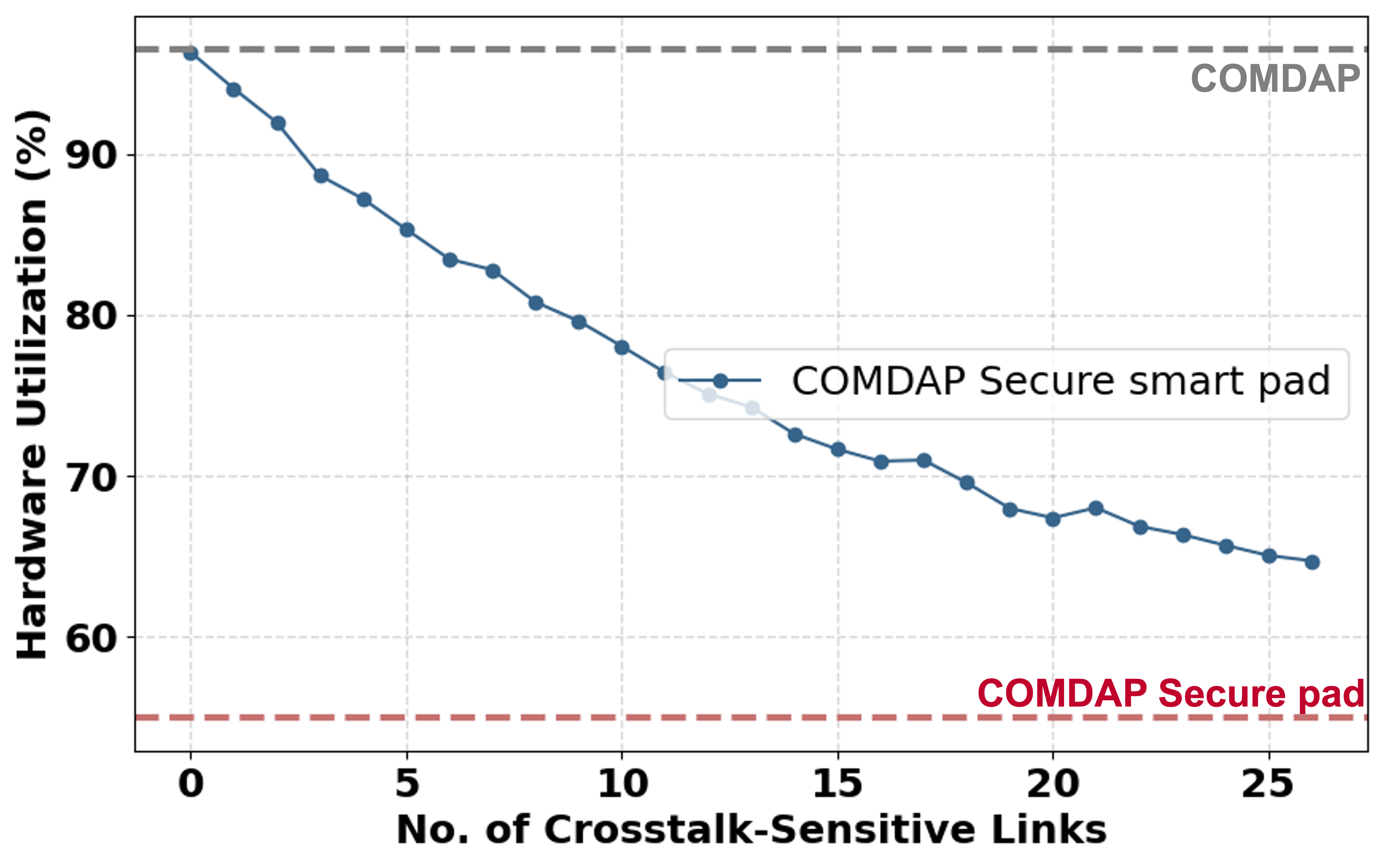}
    \caption{ Hardware utilization for COMDAP Secure smart pad with varying no. of crosstalk-prone links for Fake27QPulseV. We generate 100 different configurations for each case and report the average hardware utilization for queue\_10. }
    \label{9}
\vspace{-4mm}
\end{figure}

\subsection{Secure COMDAP}
Secure COMDAP is designed to mitigate security threats in multi-tenant quantum computing. Due to the limited availability of real hardware, we randomly designate few qubit links to be crosstalk error prone (Section IV.E). In Fig. \ref{9}, we generate 100 different configurations for each possible number of crosstalk-prone links in the hardware and report the average hardware utilization for queue\_10. We observe that Secure COMDAP, when using a general padding approach, results in a lower average hardware utilization of 55\% compared to the standard COMDAP(97\%). However, by employing a smart padding strategy, we see significant improvements in hardware utilization across all scenarios of crosstalk-prone links. We also observe an average PST reduction when compared to single execution, across queues by $\approx$4\%, and $\approx$3.5\% for the heuristics: secure COMDAP general pad and smart pad respectively (Fig. \ref{8}). We compare data across ten distinct queues, ensuring that only programs with an allocated partition shared by all three of the COMDAP heuristics were included in the plot to avoid data bias. \textit{Secure COMDAP maintains similar program performance levels as COMDAP but with lower hardware utilization.} It provides security against the following two prominent threats:

\subsubsection{Crosstalk Induced Faults} Secure COMDAP identifies and monitors qubit pairs with high error rates when operating concurrently using SRB \cite{b19}. By implementing general/smart padding, it places a protective buffer around highly susceptible qubit pairs, preventing their simultaneous operation with other pairs. It effectively reduces crosstalk and also safeguards against any adversarial programs that might exploit crosstalk to induce faults.

\subsubsection{Adversarial SWAP Injection} Previously proposed program allocation heuristics \cite{b6}\cite{b7}\cite{b8} prioritize programs with a high number of CNOT gates to allocate better-connected qubits. However, COMDAP aims to distribute resources equitably, without preferentially allocating programs based on their CNOT gate count. Our strategy effectively complicates any adversary's attempt to target specific qubits and initiate a SWAP injection attack. Despite running all programs concurrently, the average increase in ($\Delta$CX) is minimal, about 0.05X. 

\begin{figure}
    \centering
    \includegraphics[width= 3.5in]{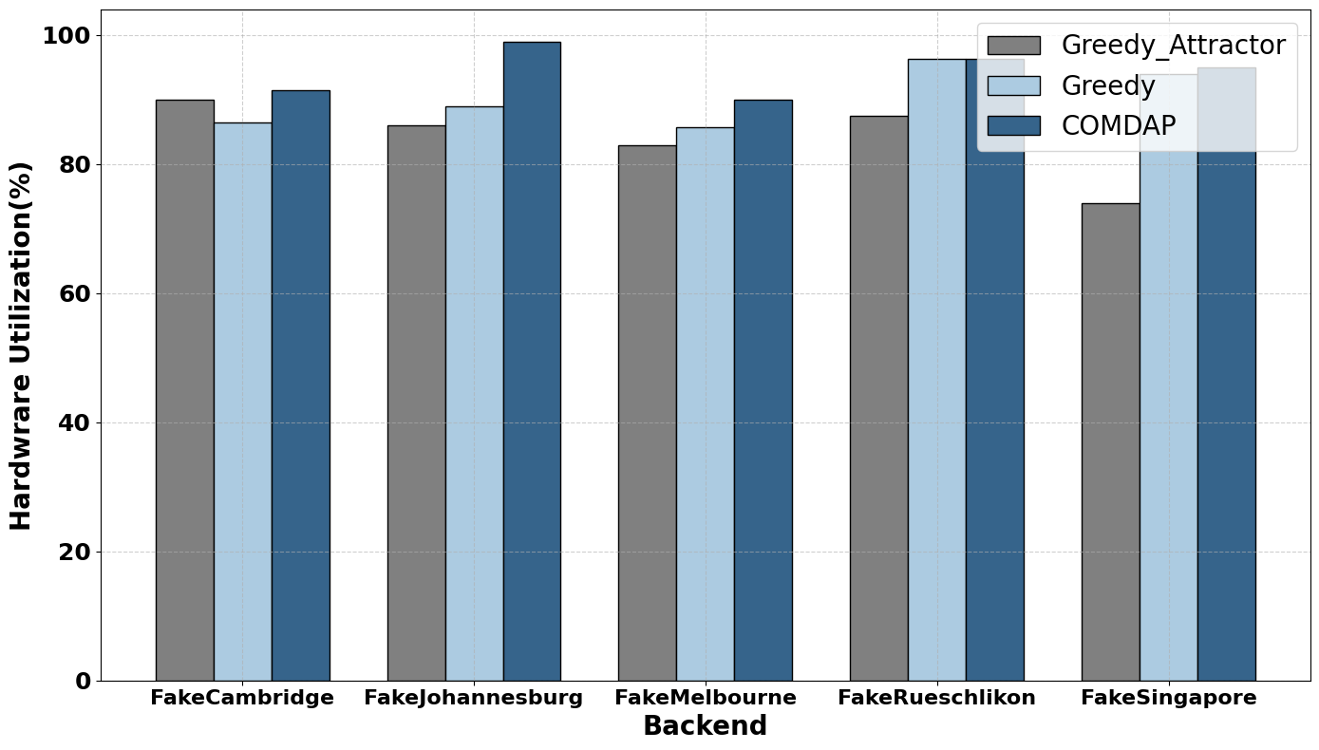}
    \caption{Hardware utilization for proposed partitioning heuristics. }
    \label{10}
\vspace{-4mm}
\end{figure}

\section{Discussion}

\subsection{Efficacy across different hardware topologies}
We test our heuristics across multiple fake provider modules, each with different configurations and sizes. Fig. \ref{10} illustrates hardware utilization for various partitioning heuristics and shows the average utilization for different 10 queues for each backend. For the attractor node-based greedy strategy, we observe the lowest hardware utilization with an average of 82\%. The COMDAP heuristic is the most effective across different hardware topologies, with an overall average of 96\%.

\subsection{Implementation on real hardware}
Fake backends, which run on classical computers, offer the advantage of quick simulations by avoiding the long queues and limited access time typical of real hardware. While benchmarks on actual hardware might yield more accurate results due to factors like crosstalk, the use of proposed heuristics—which enhance hardware utilization —would not significantly alter the overall conclusions as our heuristics account for crosstalk. One could feed the accurate crosstalk prone links to secure COMDAP by performing SRB experiments to recalibrate the partitioning and allocation. 


\section{Conclusion}
In this work, we address fundamental challenges in multi-tenant quantum computing, focusing on equitable hardware allocation, crosstalk mitigation, and security against fault injection attacks and adversarial SWAP injection attacks. We propose a novel framework that includes Community-Based Dynamic Allocation Partitioning (COMDAP) and its secure enhancement. We report a notable enhancement in resource allocation efficiency, with COMDAP achieving a 23\% higher hardware utilization rate than traditional greedy heuristics, while providing a secure multi-tenant environment.


\begin{thebibliography}{00}
\bibitem{b1} I. Cong, S. Choi, and M. D. Lukin, “Quantum convolutional neural networks,” Nature Physics, 2019. [Online]. Available: https: //doi.org/10.1038/s41567- 019- 0648- 8.
\bibitem{b2} Upadhyay, Suryansh, and Swaroop Ghosh. "Robust and Secure Hybrid Quantum-Classical Computation on Untrusted Cloud-Based Quantum Hardware." arXiv preprint arXiv:2209.11872 (2022).
\bibitem{b3} Y. Cao, J. Romero, and A. Aspuru-Guzik, “Potential of quantum computing for drug discovery,” IBM Journal of Research and Development, vol. 62, no. 6, pp. 6–1, 2018.
\bibitem{b4}E. Farhi, J. Goldstone, and S. Gutmann, “A quantum approximate optimization algorithm,” arXiv preprint arXiv:1411.4028, 2014.
\bibitem{b5}  Harper, Benjamin, et al. "Crosstalk Attacks and Defence in a Shared Quantum Computing Environment." arXiv preprint arXiv:2402.02753 (2024).
\bibitem{b6} Niu, Siyuan, and Aida Todri-Sanial. "Enabling Multi-tenant computing  mechanism for quantum computing in the NISQ era." Quantum 7 (2023): 925.
\bibitem{b7} Das, Poulami, et al. "A case for Multi-tenant computing  quantum computers." Proceedings of the 52nd Annual IEEE/ACM International Symposium on Microarchitecture. 2019.
\bibitem{b8} Liu, Lei, and Xinglei Dou. "QuCloud+: A Holistic Qubit Mapping Scheme for Single/Multi-programming on 2D/3D NISQ Quantum Computers." ACM Transactions on Architecture and Code Optimization 21.1 (2024): 1-27.
\bibitem{b9} S. Upadhyay and S. Ghosh, "Stealthy SWAPs: Adversarial SWAP Injection in Multi-Tenant Quantum Computing," 2024 37th International Conference on VLSI Design and 2024 23rd International Conference on Embedded Systems (VLSID), Kolkata, India, 2024, pp. 474-479, doi: 10.1109/VLSID60093.2024.00085.
\bibitem{b10} Blondel, Vincent D., et al. "Fast unfolding of communities in large networks." Journal of statistical mechanics: theory and experiment 2008.10 (2008): P10008.
\bibitem{b11}Iverson, Joseph K., and John Preskill. "Coherence in logical quantum channels." New Journal of Physics 22.7 (2020): 073066.
\bibitem{b12}Magesan, Easwar, et al. "Efficient measurement of quantum gate error by interleaved randomized benchmarking." Physical review letters 109.8 (2012): 080505.
\bibitem{b13}Busch, Paul, Pekka Lahti, and Reinhard F. Werner. "Colloquium: Quantum root-mean-square error and measurement uncertainty relations." Reviews of Modern Physics 86.4 (2014): 1261
\bibitem{b14} Ash-Saki, Abdullah, Mahabubul Alam, and Swaroop Ghosh. "Analysis of crosstalk in NISQ devices and security implications in multi-programming regime." In Proceedings of the ACM/IEEE International Symposium on Low Power Electronics and Design, pp. 25-30. 2020.
\bibitem{b15} Upadhyay, Suryansh, and Swaroop Ghosh. "Obfuscating quantum hybrid-classical algorithms for security and privacy." arXiv preprint arXiv:2305.02379 (2023).
\bibitem{b16} Prakash Murali, Jonathan M Baker, Ali Javadi-Abhari, Frederic T Chong, and Margaret Martonosi. 2019. Noise-adaptive compiler mappings for noisy intermediatescale quantum computers. In Proceedings of the Twenty-Fourth International Conference on Architectural Support for Programming Languages and Operating Systems. ACM, 1015–1029.
\bibitem{b18} https://quantum-computing.ibm.com/services/resources/docs/resources/
manage/systems/queue
\bibitem{b19} Gambetta, Jay M., et al. "Characterization of addressability by simultaneous randomized benchmarking." Physical review letters 109.24 (2012): 240504.
\bibitem{b20} Prakash Murali, David C McKay, Margaret Martonosi, and Ali Javadi-Abhari. Software
mitigation of crosstalk on noisy intermediate-scale quantum computers. In Proceedings of the Twenty-Fifth International Conference on Architectural Support for Programming Languages and Operating Systems, pages 1001–1016, 2020.
\bibitem{b21}Ang Li, Samuel Stein, Sriram Krishnamoorthy, and James Ang. "QASMBench: A Low-Level Quantum Benchmark Suite for NISQ Evaluation and Simulation." ACM Transactions on Quantum Computing (2022). DOI:10.1145/3550488, [arXiv:2005.13018].
\end{thebibliography}

\section{Acknowledgment}

This work is supported in parts by NSF (CNS-1722557, CNS-2129675, CCF-2210963, CCF-1718474, OIA-2040667, DGE-1723687, DGE-1821766, and DGE-2113839) and Intel's gifts.


\end{document}